# Using GANs for *De Novo* Protein Design Targeting Microglial IL-3R$\alpha$ to Inhibit Alzheimer's Progression

Arnav Swaroop[1]

## Abstract

IL-3 is a hemopoietic growth factor that usually targets blood cell precursors; IL-3R is a cytokine receptor that binds to IL-3. However, IL-3 takes on a different role in the context of glial cells in the nervous system, where studies show that the protein IL-3 protects against Alzheimer's disease by activating microglia at their IL-3R receptors, causing the microglia to clear out the tangles caused by the build-up of misfolded Tau proteins. In this study, we seek to ascertain what role the secondary structure of IL-3 plays in its binding with the receptor. The motivation behind this study is to learn more about the mechanism and identify possible drugs that might be able to activate it, in hopes of inhibiting the spread of Alzheimer's Disease. From a preliminary analysis of complexes containing IL-3 and IL-3R, we hypothesized that the binding is largely due to the interactions of three alpha helix structures stretching towards the active site on the receptor. The original Il-3 protein serves as the control in this experiment; the other proteins being tested are generated through several types of computational *de novo* protein design, where machine learning allows for the production of entirely novel structures. The efficacy of the generated proteins is assessed through docking simulations with the IL-3R receptor, and the binding poses are also qualitatively examined to gain insight into the function of the binding. From the docking data and poses, the most successful proteins were those with similar secondary structure to IL-3.

[1] The Harker School, San Jose, CA, USA

# Introduction

The goal of this research is to investigate the role the secondary structure of IL-3, specifically the alpha helices, plays in its binding with IL-3R$\alpha$ and how this can help inform future drug discovery for inhibiting Alzheimer's disease. Preliminary computational analysis of complexes containing IL-3 and IL-3R$\alpha$ shows that the binding is largely due to the interactions of three alpha helix structures, leading to the prediction that the most effective generated proteins would have alpha helices stretching towards the active site on the receptor to echo the most useful structures of the original protein.

## Alzheimer's Disease

Alzheimer's is the most common type of dementia in humans. It is a progressive disease beginning with mild memory loss and possibly leading to loss of ability to carry on simple tasks. In 2020, ~5.8 million Americans were living with Alzheimer's disease [1] For every 5 years beyond the age of 65, risk of Alzheimer's disease doubles [1]. Alzheimer's disease is the 6th leading cause of death for adults in the United States and 5th most common for seniors [1] and it is only getting more prevalent as can be seen in Figure 1.

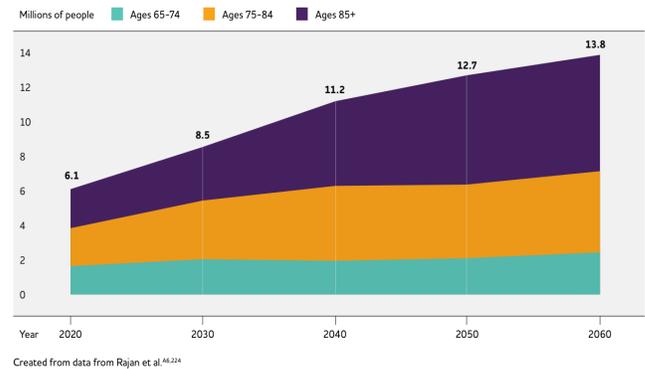

**Figure 1:** Graph from the Alzheimer's association [2] plotting predicted number of Alzheimer's patients in the US over time.

Alzheimer's disease involves parts of the brain that control thought, memory, and language. Researchers currently believe that Alzheimer's disease is primarily driven by the clustering and building-up of misfolded Tau proteins (Neurofibrillary tangles) which eventually damage neural function. Amyloid Beta plaques are also hypothesized to build up but there is some controversy about Amyloid Beta's role in Alzheimer's due to unreliable data [3]. Current pharmaceutical drugs have only temporary effects and do not slow progression of the disease: there is currently no "cure".

## The Interleukin-3 Pathway to Inhibiting Alzheimer's

Interleukin-3 (IL-3) is a hemopoietic (blood cell-related) colony stimulating factor and cytokine protein that targets blood cell precursors such as pluripotent stem cells [4]. IL-3R$\alpha$ is a cell-surface cytokine receptor that binds with IL-3. In the nervous system, IL-3 inhibits Alzheimer's disease by activating microglia (immune cells of the nervous system) at their IL-3R$\alpha$ receptors to clear Tau protein and Amyloid Beta protein plaques [5] (see Figure 2 for mechanism). IL-3 is typically released by astrocytes (glial cells that support neurons)

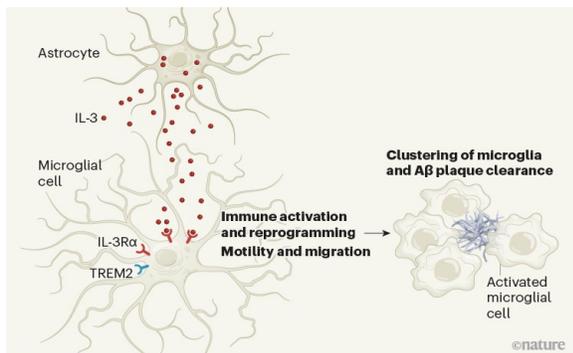

**Figure 2:** Visualization of Il-3 pathway to activating microglia [6] Studies done on mice found that the protein IL-3 effectively protects against Alzheimer's disease, both in preventing its formation and inhibiting its spread [6].

## Materials

We used the RCSB Protein Database to obtain crystallographic models of the proteins. We then used PyMol [7] for visualization and protein processing as in Figure 3. We also used Schrodinger [8] for visualizing and analyzing structures as in Figure 3. PyRosetta [10] is a python library for computational biology based on Rosetta [9] that was used for multiple purposes such as processing the protein. RamaNet [11] is a Neural Network consisting of a GAN connected to a LSTM for Helical Protein generation. We also used RosettaDesign by UNC [12] to redesign the protein based on their backbone. For testing, we used ZDOCK [13], a docking server and scoring algorithm for unbound docking using the Fast Fourier Transform algorithm. For training the model and implementing it, we used Tensorflow [14], Google's machine learning library for python. As for hardware used in the machine learning, we utilized a linux machine with an x86 CPU architecture and a Nvidia RTX Series GPU.

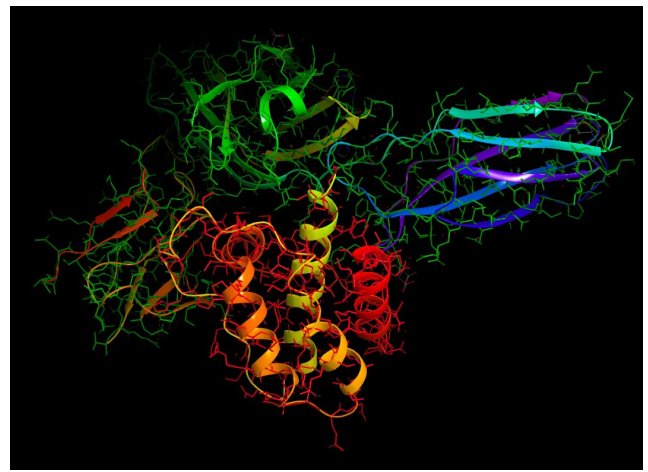

**Figure 3:** IL-3 [15] in PyMol [7] highlighting secondary structure (alpha helices) in red

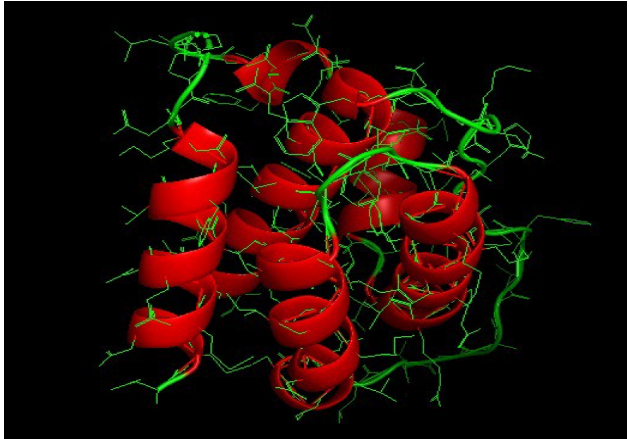

**Figure 4:** IL-3 complexed with IL-3R$\alpha$ [15] in Schrodinger Maestro [8] (red chemical structure is IL-3, green is receptor)

# Methods

Our approach is summarized with the following flowchart below (Figure 5). More details are below.

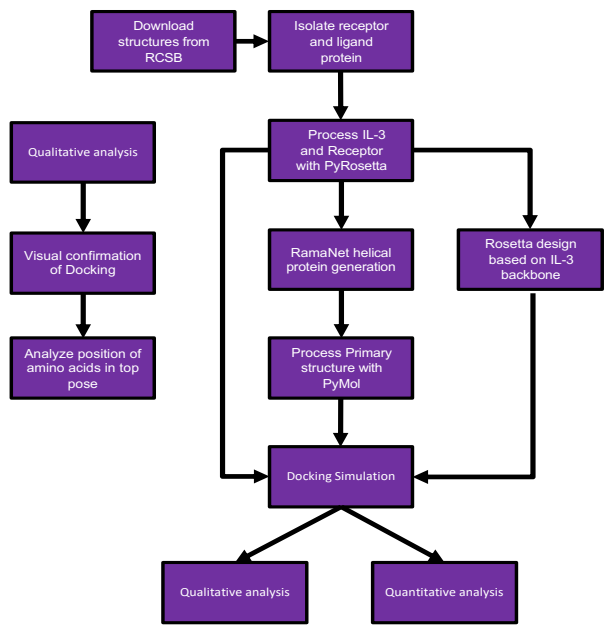

**Figure 5:** Flowchart of overall process and methods of qualitative analysis

## *De Novo* Protein Design

*De Novo* protein design is a technique where novel proteins are generated, either computationally (usually using machine learning), or manually. For this research, there were two types of *De Novo* protein design involved: helical protein-based protein design and backbone modification-based protein design. For the helical protein-based protein design, using the RamaNet neural network, we generated 10 novel helical protein structures (designated Rama 1 to Rama 10) using IL-3 as the initial backbone (see Figure 6). The advantage of this is generating completely novel structures. The drawbacks, however, are that some awkwardly shaped structures are created, and backbones had to be manually adjusted through PyMol mutagenesis. For the protein design using backbone-based modification, using RosettaDesign, we generated 10 novel proteins (designated RD1 to RD10) created by modifying the backbone (primary structure) of IL-3. These had the advantage of being able to be designed to maximize docking based off of the receptor but on the other hand, changes were mostly minimal as can be seen in Figure 7.

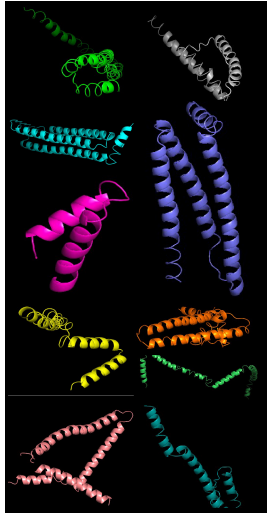

**Figure 6:** Proteins with ribbons (secondary structure) generated by RamaNet in PyMol [7]

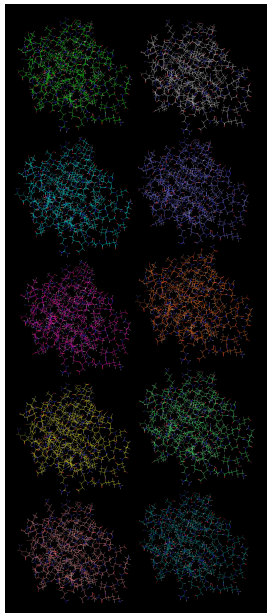

**Figure 7:** Proteins generated by RosettaDesign with primary structure in PyMol [7]

## Docking simulations

We used ZDOCK for docking simulations to test the effectiveness of the generated proteins and ran one ZDOCK simulation between each of the 20 generated proteins and the original IL-3 protein (the control) as the ligand protein with the IL-3Ra receptor. Each simulation had 2000 different poses and we used the highest scoring case for data analysis for accurate measurement of efficacy. However, qualitative analysis was still necessary to analyze the effects of secondary and primary structure.

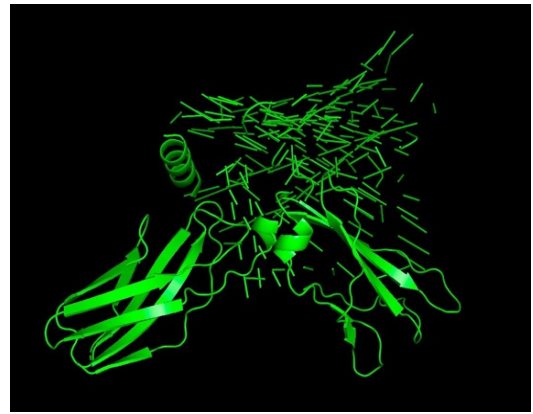

**Figure 8:** ZDOCK output of RD 1 and IL- 3Ra [15] in PyMol [7]

## Results and Discussion

a)

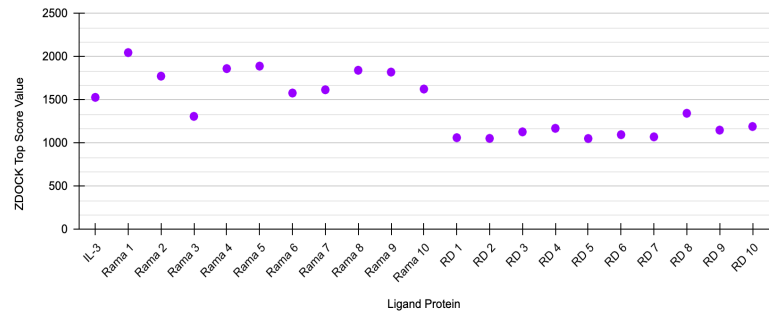

b)

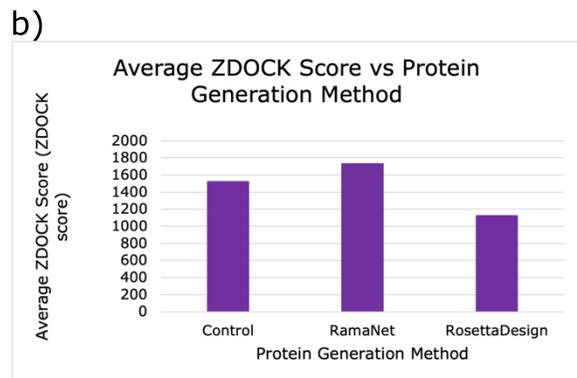

c)

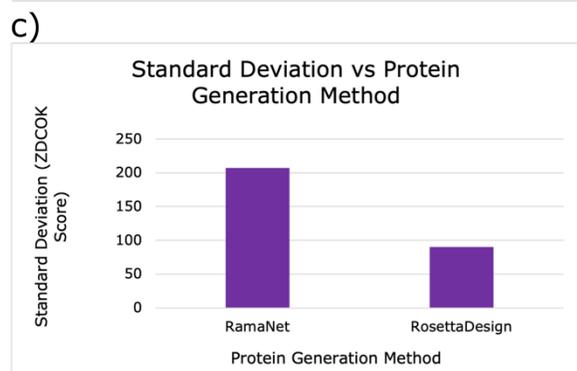

**Figure 9:** a) Effectiveness of generated protein in docking, b) average effectiveness of protein by method, c) standard deviation of effectiveness

The helical proteins seemed to vastly outperform the ones generated by backbone modification. The helical proteins also outperformed the control by a large margin, showing the potential for novel helical proteins as potential drugs for Alzheimer's. Of the helical proteins, the ones that performed best were Rama 1, Rama 4, and Rama 5. All had multiple parallel alpha helices that binded like the original IL-3 protein at the correct site for the best case (see Figure 8). On average, the helical proteins were significantly larger. Some were even larger than the receptor. Since the receptor is kept stationary during docking simulations, this may have contributed to the much higher docking score of the helical proteins.

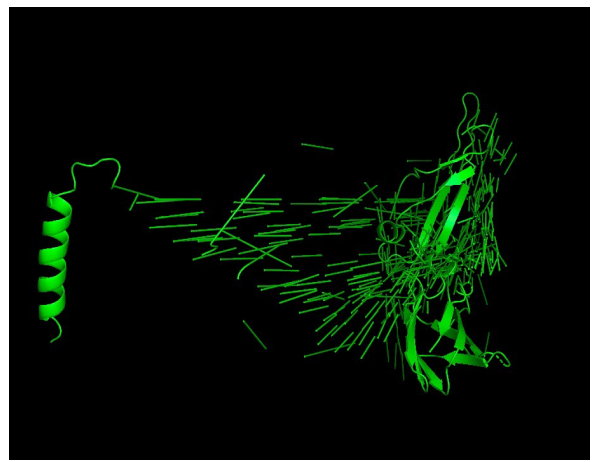

**Figure 10:** ZDOCK output of RM 1 and IL-3R$\alpha$ [15] in PyMol [7]

From qualitative analysis, helical proteins' docking often distorted both the ligand and receptor protein by much more than for the backbone-modification proteins (compare Figure 10 to Figure 8 and then to Figure 4 for the original receptor complex).
We noticed that backbone-modification proteins that performed worst had different amino acids near the docking site. The backbone-modification proteins had a much lower standard deviation than the helical proteins, suggesting that the positioning of the alpha helices is very important.

# Conclusions

We found that the direction, position, and number of the alpha helices is very important for docking to the IL-3Ra receptor: the best performing proteins have 2-3 parallel alpha helices that intersect the active site of the receptor at nearly perpendicular angles. We also found that the backbone is somewhat important as various different types of backbone modification all resulted in lower docking scores. Thus, we believe that future drugs for Alzheimer's should either use supplements of IL-3 or novel helical proteins like Rama 1, Rama 4, and Rama 5. The novel helical proteins have the benefit of not causing damage by binding and activating other proteins in the brain and in the body as IL-3 has other functions in the body.

# Future work

In the future, we would like to run different kinds on docking simulations (PIPER, Gramm, etc.) for deeper qualitative analysis. With our funds and resources available, we were not able to use these other tools in this work.
We would also like to consider the generation of De Novo ligands as these could be effective and cheaper to produce.
Also, we would like to optimize this approach using other advanced machine learning models and create a GAN dedicated for specifically generating proteins similar to IL-3.
Lastly, we would like to compare our computational data against clinical data by producing and testing these proteins *in vitro* and then hopefully *in vivo*.

# Bibliography

1. Centers for Disease Control and Prevention. (2020, October 26). *What is alzheimer's disease?* Centers for Disease Control and Prevention. Retrieved March 6, 2023, from https://www.cdc.gov/aging/aginginfo/alzheimers.htm
2. Morris, G.P., Clark, I.A. & Vissel, B. Inconsistencies and Controversies Surrounding the Amyloid Hypothesis of Alzheimer's Disease. *acta neuropathol commun* 2, 135 (2014). https://doi.org/10.1186/s40478-014-0135-5
3. *Alzheimer's disease facts and figures*. Alzheimer's Disease and Dementia. (2022). Retrieved March 6, 2023, from https://www.alz.org/alzheimers-dementia/facts-figures
4. Alberts, B., Gray, D., Lewis, J., Watson, J. D., Roberts, K., & Raff, M. (1994). *Molecular Biology of the Cell* (3rd ed.). Garland Science.
5. McAlpine, C. S., Park, J., Griciuc, A., Kim, E., Choi, S. H., Iwamoto, Y., Kiss, M. G., Christie, K. A., Vinegoni, C., Poller, W. C., Mindur, J. E., Chan, C. T., He, S., Janssen, H., Wong, L. P., Downey, J., Singh, S., Anzai, A., Kahles, F., … Swirski, F. K. (2021, July 14). *Astrocytic interleukin-3 programs microglia and limits alzheimer's disease*. Nature News. Retrieved October 14, 2022, from https://www.nature.com/articles/s41586-021-03734-6#Sec1
1. Barron, J. J., & Molofsky, A. V. (2021, July 14). *A protective signal between the brain's supporting cells in alzheimer's disease*. Nature News. Retrieved October 14, 2022, from https://www.nature.com/articles/d41586-021- 01870-7
2. The PyMOL Molecular Graphics System, Version 2.0 Schrödinger, LLC.
3. Schrödinger Release 2022-3: Maestro, Schrödinger, LLC, New York, NY, 2021.
4. *Welcome to rosettacommons*. RosettaCommons. (n.d.). Retrieved October 14, 2022, from https://www.rosettacommons.org/
5. *Pyrosetta*. PyRosetta. (n.d.). Retrieved October 14, 2022, from https://www.pyrosetta.org/
6. Sabban, S., & Markovsky, M. (2020). Ramanet: Computational de novo helical protein backbone design using a long short-term memory generative adversarial neural network. *F1000Research*, *9*, 298. https://doi.org/10.12688/f1000research.22907.1
7. Drew, K., Renfrew, P. D., Craven, T. W., Butterfoss, G. L., Chou, F. C., Lyskov, S., Bullock, B. N., Watkins, A., Labonte, J. W., Pacella, M., Kilambi, K. P., Leaver-Fay, A., Kuhlman, B., Gray, J. J., Bradley, P., Kirshenbaum, K., Arora, P. S., Das, R., & Bonneau, R. (2013). Adding diverse noncanonical backbones to rosetta: enabling peptidomimetic design. *PloS one*, *8*(7), e67051. https://doi.org/10.1371/journal.pone.0067051

8. Pierce, B. G., Wiehe, K., Hwang, H., Kim, B. H., Vreven, T., & Weng, Z. (2014). ZDOCK server: interactive docking prediction of protein-protein complexes and symmetric multimers. *Bioinformatics (Oxford, England)*, *30*(12), 1771–1773. https://doi.org/10.1093/bioinformatics/btu097
9. Abadi, Mart'in, Barham, P., Chen, J., Chen, Z., Davis, A., Dean, J., … others. (2016). Tensorflow: A system for large-scale machine learning. In 12th $USENIX$ Symposium on Operating Systems Design and Implementation ($OSDI$ 16) (pp. 265–283).
10. Broughton, S.E., Hercus, T.R., Nero, T.L. et al. A dual role for the N-terminal domain of the IL-3 receptor in cell signalling. Nat Commun 9, 386 (2018). https://doi.org/10.1038/s41467-017-02633-7